\documentclass[twocolumn]{webofc}

\usepackage[varg]{txfonts}
\usepackage{physics}
\usepackage{hyperref}

\usepackage{lipsum} 

\begin{document}
\title{A Problem in the Statistical Description of Beta-Delayed Neutron Emission}
\author{\firstname{Oliver} \lastname{Gorton}\inst{1,2}\fnsep\thanks{\email{ogorton@sdsu.edu}} \and
        \firstname{Calvin} \lastname{Johnson}\inst{1}\fnsep\thanks{\email{cjohnson@sdsu.edu}} \and
        \firstname{Jutta} \lastname{Escher}\inst{2}\fnsep\thanks{\email{escher1@llnl.gov}}
}
\institute{San Diego State University, California, USA
\and
           Lawrence Livermore National Laboratory, California, USA
}
%
%
\abstract{Reaction measurements on fission products are being planned at both Argonne National Lab and at the Facility for Rare Isotope Beams. These indirect experiments produce specific short-lived nuclei via beta decay, and the subsequent neutron and gamma emission are studied. Some initial experiments found a surprising overabundance of gamma emission, which theory has yet to explain. To remedy this, we are developing an integrated nuclear data workflow that connects advanced nuclear shell model codes for describing the beta decay with a contemporary nuclear reaction model code.}
\maketitle
\begin{figure}[ht]
    \centering
    \includegraphics[width=0.4\textwidth]{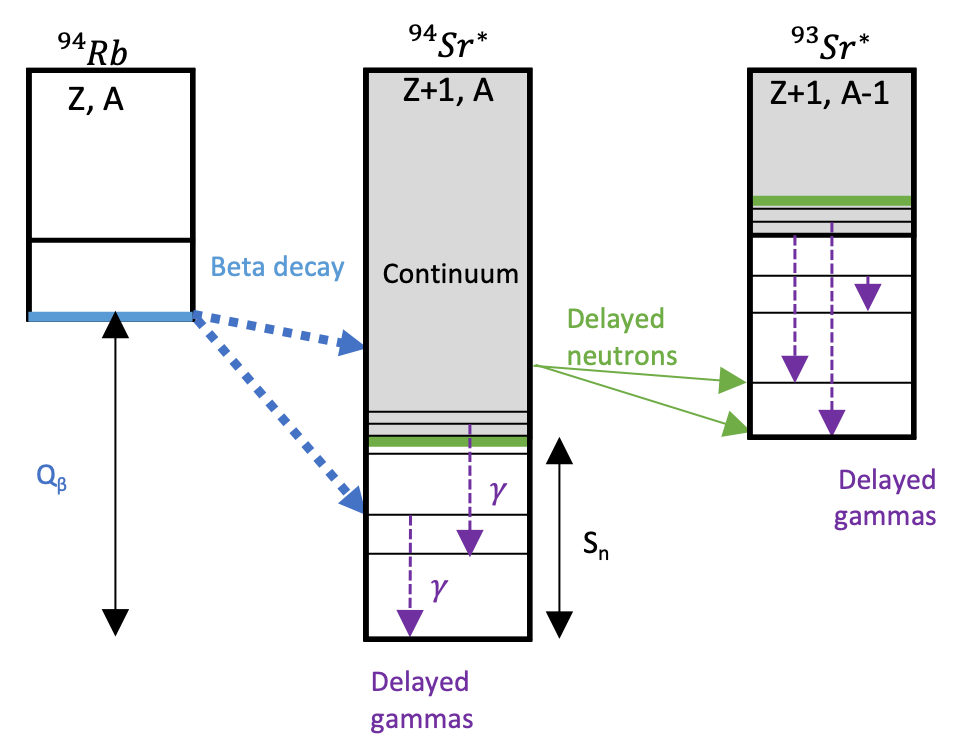}
    \caption{Schematic depiction of beta-delayed neutron emission for the case of $^{94}$Rb. The distribution of states populated by beta decay can be predicted with a microscopic structure model like quasiparticle random-phase approximation or shell model. The decay of the neutron and gamma emitting nucleus ($^{94}$Sr) is historically described by a statistical model, assuming compound nuclear decay. }
    \label{fig:bdne}
\end{figure}

Beta decay is the  mechanism for element transmutation towards stability, and plays an important role in competition with neutron-capture in the formation of heavy elements: Understanding this is one of the central tasks of the nuclear theory community and the new Facility for Rare Isotope Beams (FRIB) \cite{ARCONES20171,https://doi.org/10.48550/arxiv.2205.07996}. A less common, yet important \cite{MUMPOWER201686} process is the emission of one or more neutrons immediately following beta decay in a process called beta-delayed neutron emission (BDNE), depicted in Figure \ref{fig:bdne}. 

\section{Statistical description of BDNE does not explain experiments}
Theoretical description of BDNE has been an ongoing challenge since the late 1970's \cite{PhysRevLett.33.859}. 
The idea of \textit{selectivity}\cite{PhysRevLett.33.859} of states populated by beta decay, leading to non-statistical neutron emission, was proposed to explain peak-features seen in delayed neutron spectra. These effects were almost explained away by the \textit{Pandemonium effect} \cite{HARDY1977307}, which indicated that such peaks were artifacts due to detector limitations. Others showed that in some cases, (a) statistical models of BDNE could reproduce the peaks if excited states in the residual nucleus were included \cite{PRUSSIN1979396}, and (b) that states populated in beta decay could be strongly connected to excited states in the final nucleus \cite{SHIHABELDIN1977143}, supporting selectivity. At the close of the decade, the nuances were better understood, including the importance beta-decay strength function shape and nuclear level densities \cite{HOFF19819,PhysRevC.28.602}.

With the development of total absorption gamma-ray spectroscopy (TAGS) \cite{ALGORA1999727c,PhysRevC.60.024315} and its application to the study of BDNE \cite{PhysRevLett.115.062502, PhysRevC.95.024320, PhysRevC.100.044305}, the Pandemonium effect can be avoided. Even so, the statistical description in some cases significantly under-predicts the intensity of emitted photons from TAGS experiments, e.g. in \cite{PhysRevC.95.024320}. Thus, BDNE remains an important and unresolved modeling challenge.
There are three hypotheses which may explain this.

\textit{Hyp. 1: Photon-decay strength function is stronger than assumed.} This is the simplest explanation: if the photon-strength function is enhanced, the nucleus formed by beta decay will de-excite below the neutron separation energy before delayed neutrons are emitted. If this is the explanation, then many neutron capture rates will also have to be re-evaluated, especially for astrophysical interests.

\textit{Hyp. 2: Forbidden beta decay is stronger than assumed, blocking high-$l$ neutron emission.} If forbidden transitions play a significant role in these neutron-rich cases, then higher angular momentum states populated by beta decay would block this emission of neutrons to available states in the neighboring nucleus.

\textit{Hyp. 3: Beta decay does not lead to a compound nucleus.} Essentially, selectivity. This explanation requires the greatest change in our description of beta-delayed neutron emission. It would also have significant implications for the use of beta-decay as a means of indirect cross section measurements such as the beta-Oslo method \cite{PhysRevLett.113.232502}.

\section{Applying modern methods in shell model calculations}
Today's Hauser-Feshbach (HF) codes for reactions have evolved into multi-physics packages that go beyond purely statistical \cite{PhysRev.87.366} decay models. It's now routine to have an integrated software package that computes coupled-channel optical models, pre-equilibrium decays, width-fluctuation corrections, multi-chance fission, decays between hundreds of discrete states, and more \cite{KONING20191,YAHFC:18}. 

Some BDNE studies have made progress combining quasiparticle random-phase approximation (QRPA)-type descriptions of beta decay with HF calculations to reasonable effect \cite{PhysRevC.78.054601,HALL2021136266}. We are also aware of recent efforts with the shell model \cite{grzywacz2022evidence}.
We want to continue along these lines to incorporate modern shell model methods into HF codes in a self-consistent way, including level densities and gamma-ray strength functions derived from the shell model.

\subsection{Proton and Neutron Approximate Shell Model}

Shell model calculations, treating nuclei at the nucleon-degree-of-freedom, offer the largest model spaces to capture the physics of complex nuclei. Energy levels, binding energies, nuclear level densities, gamma-ray strength functions, and beta-decay strength functions can all be computed from the same wave functions generated from phenomenological interactions. 
Even after restricting configurations to a finite valence space, nuclei of interest to BDNE generate basis dimensions several orders of magnitude larger than current computational limits, $O(10^{10})$. 

To make these models tractable without discarding important orbitals, we apply an importance truncation in the many-body configuration space.
We assign configuration importance, in explicit proton-neutron formalism based, based on the eigenvalues of the "separable" proton and neutron parts of the Hamiltonian. This approach is motivated by the empirical fact that eigenstates of the nuclear Hamiltonian $H = H^{(p)} + H^{(n)} + H^{(pn)}$ are well approximated by simple products of eigenstates of the proton-only and neutron-only interactions, $H^{(p)}$ and $H^{(n)}$ \cite{PhysRevC.67.051303, PhysRevC.69.024312}, with fidelity increasing exponentially with the number of extremal states taken in combination. 

Our code, PANASH (proton and neutron approximate shell-model, unpublished), implements this importance truncation scheme. Basis states are constructed by coupling together eigenstates of $H^{(p)}$ and $H^{(n)}$, up to fixed total angular momentum and parity. The basis is then truncated by using only some fraction of all states, taking the lowest excitations from each subspace. Similar work has been demonstrated by others \cite{PhysRevC.67.051303, PhysRevC.69.024312}. 
PANASH has been designed and simplified for our purpose, and is ready for extensive application.

We present some preliminary results computed with PANASH. These include the energy level spectra (and binding energies), beta transition strengths, photon strength functions (PSF, or gamma ray strength functions ), and nuclear level densities (NLDs). All of these can be used as input to the HF modeling of beta-delayed neutron emission.

\subsection{Spectra of a complex nucleus in the $fpg$ shell}

$^{70}$Ge is a complex nucleus, thought to exhibit triaxiality \cite{PhysRevC.76.034317}. We use this as a rigorous test of our many-body method, as has been done recently for other methods \cite{Lauber_2021}. We work in the $fpg$ model space, using the JUN45 interaction \cite{PhysRevC.80.064323}. 

In this case, we construct the basis by taking all combinations of proton and neutron components which satisfy the limit on component excitation energies: $(E^{(p)})^2+(E^{(n)})^2<\epsilon^2$, for some truncation limit $\epsilon$. This leads to a maximum model space dimension of $10^4$ at $\epsilon=8.5$ MeV, using 217 of 701 protons components and 1190 of 47722 neutron components.

\begin{figure}[ht]
    \centering
    \includegraphics[width=0.45\textwidth]{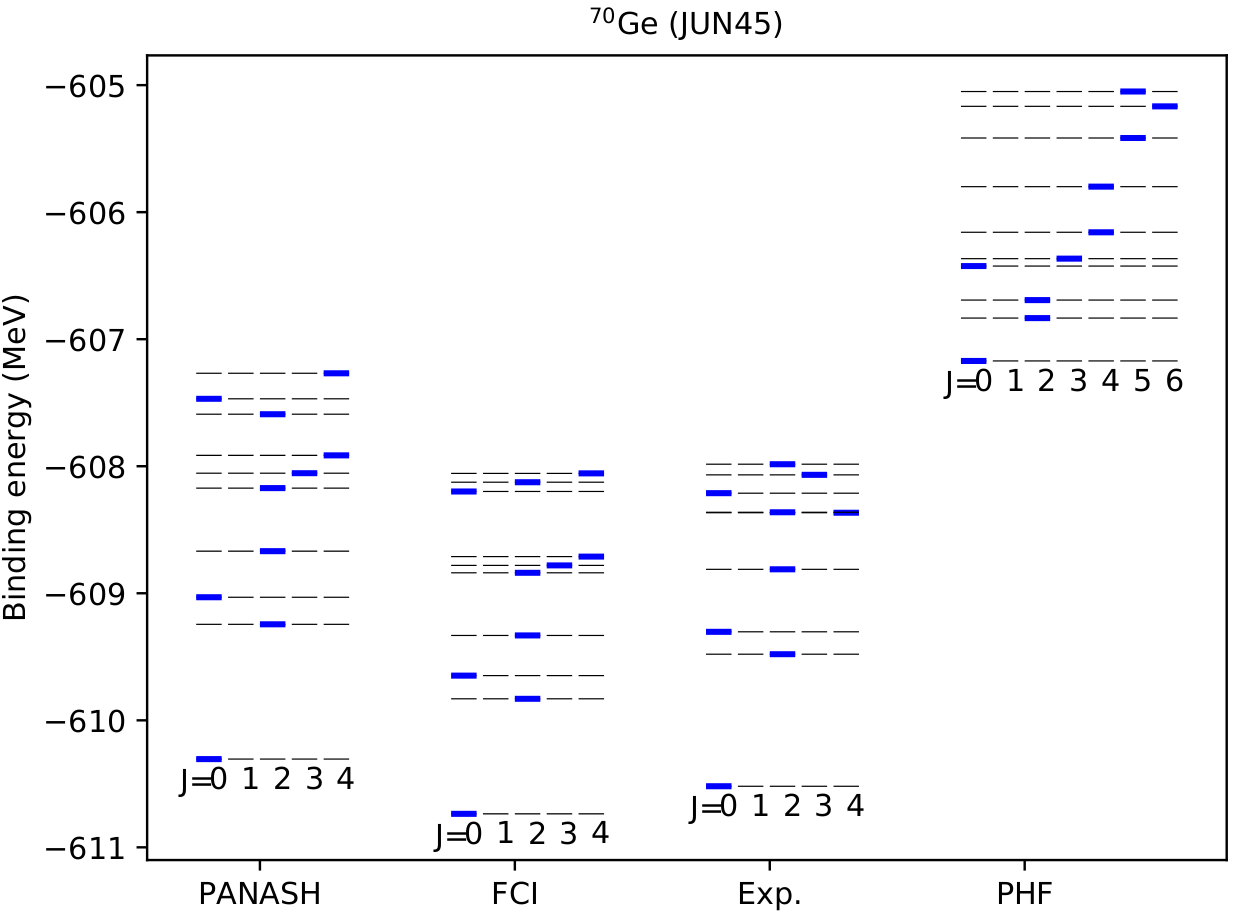}
    \caption{Low-lying binding-energy spectra for $^{70}$Ge from PANASH compared to the full configuration interaction (FCI, untruncated model), a projected Hartree-Fock calculation using the same interaction \cite{Lauber_2021}, and experimental values \cite{GURDAL20161}. In this case, the truncation is such that $(E^{(p)})^2+(E^{(n)})^2< 8.5$ MeV. }
    \label{fig: convergencee}
\end{figure}

Results are shown in Figures \ref{fig: convergencee}. The error of our ground-state binding energy is comparable to that of the full configuration interaction (FCI, untruncated model) result. It is also a significant improvement over projected Hartree-Fock (PHF) performed using the same interaction \cite{Lauber_2021}. PHF is a method where a deformed (possibly triaxially deformed) Hartree-Fock state is projected to have good angular momentum, and then used as a reference state for the random phase approximation (RPA) (which is known to be unreliable for reproducing binding energies \cite{PhysRevC.66.034301}). We also successfully reproduce the third $0^+$ state, which is missing from the PHF calculation. Overall this is evidence that PANASH is able to capture important many-body correlations.

\subsection{Gamow-Teller decay}

\begin{figure}[ht]
    \centering
    \includegraphics[width=0.45\textwidth]{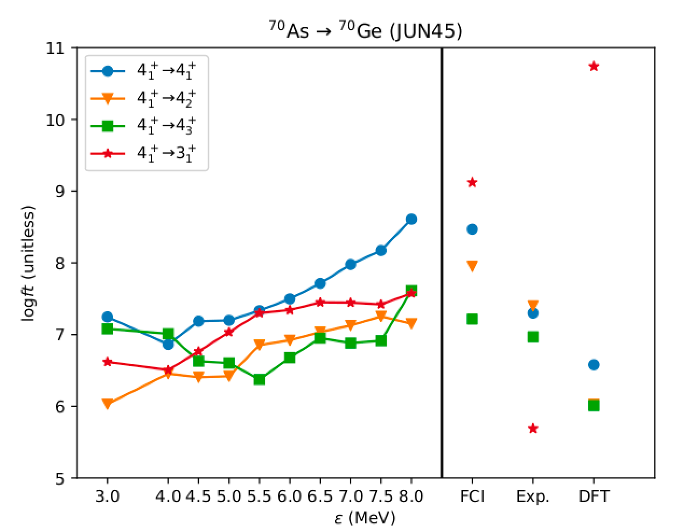}
    \caption{Decay half-lives given by log$ft$s from the ground state of $^{70}$As, as a function of the maximum excitation energy $\epsilon$ of subspace states used to construction the basis. Comparison to DFT results \cite{PhysRevC.105.044306} shown. }
    \label{fig: convergencebgt}
\end{figure}

We compute beta-decay log$ft$ values for the decay of $^{70}$As to $^{70}$Ge. We use the same model space and interaction as in the previous section, and a Gamow-Teller quenching factor of $q=0.684$, taken from \cite{Kumar_2016}.
We also use the same basis truncation method as before, but this time we track the evolution of the results as a function of the truncation parameter $\epsilon$. 
Results are shown in Figure \ref{fig: convergencebgt}.
We find that the error of beta decay $\log ft$ values are comparable to FCI error relative to experiment.

\subsection{GSFs and NLDs}

For demonstration of PSF's and NLDs, we show preliminary results for $^{78}$Ge in the pf-shell ($f_{5/2},p_{3/2},p_{1/2},g_{9/2}$ valence orbitals above a $^{56}$Ni core) with the JUN45 interaction \cite{PhysRevC.80.064323}. 
In this case, we take 20 percent of the proton and neutron eigenstates to construct the basis, yielding a maximum basis dimension 7,000, only 4-percent relative to the un-truncated dimension of 170,000. 
(This is the dimension of the largest $J$-basis, $J=6$.) 
To calculate the M1 strength functions involved in the PSF,
we use the standard magnetic moment operator with proton spin and orbital couplings
$g_s=5.586$, $g_l=1$; and neutron couplings $g_s=-3.826$, $g_l=0$.
We used the usual quenching factor of $q=0.7$ \cite{PhysRevC.80.064323}.

\begin{figure}[ht]
    \centering
    \includegraphics[width=0.5\textwidth]{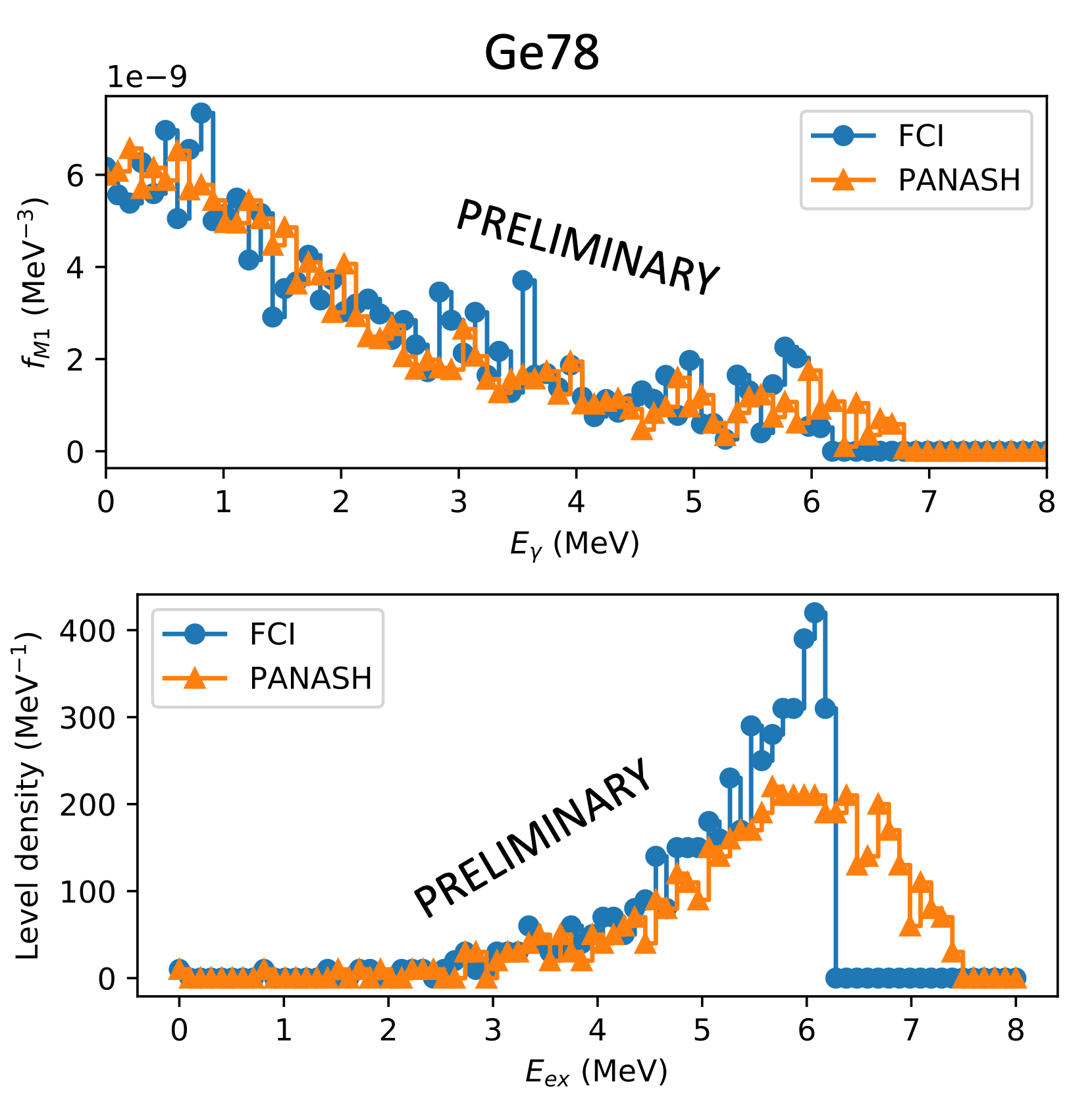}
    \caption{Preliminary calculations of photon strength functions (PSFs) (top panel) and nuclear level densities (NLDs) (bottom panel) using the PANASH code, with comparison to the full configuration interaction (FCI), untruncated model. In both cases, the first 500 positive-parity states were computed. Energy bins are 0.1 MeV.}
    \label{figgsfld}
\end{figure}

We compute the first 500 positive-parity states for use in the calculation, for $J=0-10$. The resulting level density and M1 photon strength function is shown in Figure \ref{figgsfld}. We find good agreement with recent results from \cite{PhysRevC.105.034335}, which used a similar interaction with the same single particle space. Although that calculation also included negative parity states, we find that these only slightly alter the PSF.

\section{Going forward}

We estimate that the case of $^{94}$Rb, relevant to \cite{PhysRevC.95.024320}, would have a dimension $O(10^{13})$ in an untruncated FCI calculation. By contrast, the proton/neutron subspaces are $O(10^{7})$, which can be performed on a modest workstation. A basis constructed with these components could then be extended as far as achievable on available resources to maximize fidelity. A major remaining issue will be the development of effective interactions for previously unreachable model spaces.

\section{Acknowledgements}

We thank Arnd Junghans and Ronald Schwengner for feedback on an early version of this manuscript, and Robert Grzywacz for helpful discussions.

This work was performed under the auspices of the U.S. Department of Energy by Lawrence Livermore National Laboratory under Contract DE-AC52-07NA27344, with support from the ACT-UP award.

\bibliography{gortonbib}

\begin{thebibliography}{30}

\bibitem{ARCONES20171}
A.~Arcones, D.W. Bardayan, T.C. Beers, L.A. Bernstein, J.C. Blackmon,
  B.~Messer, B.A. Brown, E.F. Brown, C.R. Brune, A.E. Champagne et~al.,
  Progress in Particle and Nuclear Physics \textbf{94}, 1 (2017)

\bibitem{https://doi.org/10.48550/arxiv.2205.07996}
H.~Schatz, A.D.B. Reyes, A.~Best, E.F. Brown, K.~Chatziioannou, K.A. Chipps,
  C.M. Deibel, R.~Ezzeddine, D.K. Galloway, C.J. Hansen et~al., \emph{Horizons:
  Nuclear astrophysics in the 2020s and beyond} (2022),
  \urlstyle{tt}\url{https://arxiv.org/abs/2205.07996}

\bibitem{MUMPOWER201686}
M.~Mumpower, R.~Surman, G.~McLaughlin, A.~Aprahamian, Progress in Particle and
  Nuclear Physics \textbf{86}, 86 (2016)

\bibitem{PhysRevLett.33.859}
H.~Franz, J.V. Kratz, K.L. Kratz, W.~Rudolph, G.~Herrmann, F.M. Nuh, S.G.
  Prussin, A.A. Shihab-Eldin, Phys. Rev. Lett. \textbf{33}, 859 (1974)

\bibitem{HARDY1977307}
J.~Hardy, L.~Carraz, B.~Jonson, P.~Hansen, Physics Letters B \textbf{71}, 307
  (1977)

\bibitem{PRUSSIN1979396}
S.~Prussin, Z.~Oliveira, K.L. Kratz, Nuclear Physics A \textbf{321}, 396 (1979)

\bibitem{SHIHABELDIN1977143}
A.~Shihab-Eldin, W.~Halverson, F.~Nuh, S.~Prussin, W.~Rudolph, H.~Ohm,
  K.~Kratz, Physics Letters B \textbf{69}, 143 (1977)

\bibitem{HOFF19819}
P.~Hoff, Nuclear Physics A \textbf{359}, 9 (1981)

\bibitem{PhysRevC.28.602}
S.~Raman, B.~Fogelberg, J.A. Harvey, R.L. Macklin, P.H. Stelson, A.~Schr\"oder,
  K.L. Kratz, Phys. Rev. C \textbf{28}, 602 (1983)

\bibitem{ALGORA1999727c}
A.~Algora, D.~Cano-Ott, B.~Rubio, J.~Tain, J.~Agramunt, J.~Blomqvist,
  L.~Batist, R.~Borcea, R.~Collatz, A.~Gadea et~al., Nuclear Physics A
  \textbf{654}, 727c (1999)

\bibitem{PhysRevC.60.024315}
Z.~Hu, L.~Batist, J.~Agramunt, A.~Algora, B.A. Brown, D.~Cano-Ott, R.~Collatz,
  A.~Gadea, M.~Gierlik, M.~G\'orska et~al., Phys. Rev. C \textbf{60}, 024315
  (1999)

\bibitem{PhysRevLett.115.062502}
J.L. Tain, E.~Valencia, A.~Algora, J.~Agramunt, B.~Rubio, S.~Rice, W.~Gelletly,
  P.~Regan, A.A. Zakari-Issoufou, M.~Fallot et~al., Phys. Rev. Lett.
  \textbf{115}, 062502 (2015)

\bibitem{PhysRevC.95.024320}
E.~Valencia, J.L. Tain, A.~Algora, J.~Agramunt, E.~Estevez, M.D. Jordan,
  B.~Rubio, S.~Rice, P.~Regan, W.~Gelletly et~al., Phys. Rev. C \textbf{95},
  024320 (2017)

\bibitem{PhysRevC.100.044305}
V.~Guadilla, J.L. Tain, A.~Algora, J.~Agramunt, D.~Jordan, M.~Monserrate,
  A.~Montaner-Piz\'a, E.~N\'acher, S.E.A. Orrigo, B.~Rubio et~al., Phys. Rev. C
  \textbf{100}, 044305 (2019)

\bibitem{PhysRevLett.113.232502}
A.~Spyrou, S.N. Liddick, A.C. Larsen, M.~Guttormsen, K.~Cooper, A.C. Dombos,
  D.J. Morrissey, F.~Naqvi, G.~Perdikakis, S.J. Quinn et~al., Phys. Rev. Lett.
  \textbf{113}, 232502 (2014)

\bibitem{PhysRev.87.366}
W.~Hauser, H.~Feshbach, Phys. Rev. \textbf{87}, 366 (1952)

\bibitem{KONING20191}
A.~Koning, D.~Rochman, J.C. Sublet, N.~Dzysiuk, M.~Fleming, S.~{van der Marck},
  Nuclear Data Sheets \textbf{155}, 1 (2019), special Issue on Nuclear Reaction
  Data

\bibitem{YAHFC:18}
W.E. Ormand, \emph{{YAHFC-MC}: A {M}onte {C}arlo based {H}auser-{F}eshbach
  reaction code system} (2018),
  \urlstyle{tt}\url{http://meetings.aps.org/link/BAPS.2018.HAW.MG.8}

\bibitem{PhysRevC.78.054601}
T.~Kawano, P.~M\"oller, W.B. Wilson, Phys. Rev. C \textbf{78}, 054601 (2008)

\bibitem{HALL2021136266}
O.~Hall, T.~Davinson, A.~Estrade, J.~Liu, G.~Lorusso, F.~Montes, S.~Nishimura,
  V.~Phong, P.~Woods, J.~Agramunt et~al., Physics Letters B \textbf{816},
  136266 (2021)

\bibitem{grzywacz2022evidence}
R.~Grzywacz, J.~Heideman, M.~Madurga, Z.~Xu, T.~Kawano, J.~Escher, T.~King,
  R.~Lica, Bulletin of the American Physical Society  (2022)

\bibitem{PhysRevC.67.051303}
T.~Papenbrock, D.J. Dean, Phys. Rev. C \textbf{67}, 051303 (2003)

\bibitem{PhysRevC.69.024312}
T.~Papenbrock, A.~Juodagalvis, D.J. Dean, Phys. Rev. C \textbf{69}, 024312
  (2004)

\bibitem{PhysRevC.76.034317}
L.~Guo, J.A. Maruhn, P.G. Reinhard, Phys. Rev. C \textbf{76}, 034317 (2007)

\bibitem{Lauber_2021}
S.M. Lauber, H.C. Frye, C.W. Johnson, Journal of Physics G: Nuclear and
  Particle Physics \textbf{48}, 095107 (2021)

\bibitem{PhysRevC.80.064323}
M.~Honma, T.~Otsuka, T.~Mizusaki, M.~Hjorth-Jensen, Phys. Rev. C \textbf{80},
  064323 (2009)

\bibitem{PhysRevC.66.034301}
I.~Stetcu, C.W. Johnson, Phys. Rev. C \textbf{66}, 034301 (2002)

\bibitem{GURDAL20161}
G.~Gürdal, E.~McCutchan, Nuclear Data Sheets \textbf{136}, 1 (2016)

\bibitem{Kumar_2016}
V.~Kumar, P.C. Srivastava, H.~Li, Journal of Physics G: Nuclear and Particle
  Physics \textbf{43}, 105104 (2016)

\bibitem{PhysRevC.105.044306}
K.~Nomura, Phys. Rev. C \textbf{105}, 044306 (2022)

\end{thebibliography}
\end{document}